\documentclass{PoS}

\newcommand{\half}{\mbox{\small $\frac{1}{2}$}}          
\newcommand{\third}{\mbox{\small $\frac{1}{3}$}}         
\newcommand{\twothird}{\mbox{\small $\frac{2}{3}$}}      
\newcommand{\R}{\mbox{\tiny $R$}}                        
\newcommand{\Si}{\mbox{\tiny $S$}}                       
\newcommand{\NS}{\mbox{\tiny $N\!S$}}                    

\def\lsim{\mathrel{\rlap{\lower4pt\hbox{\hskip1pt$\sim$}}
    \raise1pt\hbox{$<$}}}                
\def\gsim{\mathrel{\rlap{\lower4pt\hbox{\hskip1pt$\sim$}}
    \raise1pt\hbox{$>$}}}                

\newcommand{\mbar}{\overline{m}}
\newcommand{\muh}{\delta{m}_u}
\newcommand{\mdh}{\delta{m}_d}
\newcommand{\msh}{\delta{m}_s}

\title{
\vspace*{-1.25cm}
\begin{minipage}{\textwidth}
\begin{flushright}
\texttt{\footnotesize
PoS(Lattice 2010)122 \\%
DESY 10-222          \\%
Edinburgh 2010/30    \\%
Liverpool LTH 889    \\%
}
\end{flushright}
\end{minipage}\\[15pt]
\vspace*{+1.25cm}
       Flavour symmetry breaking and tuning the strange quark mass  
       for 2+1 quark flavours}

\ShortTitle{Flavour symmetry breaking $\ldots$ 
            \hspace*{2.53in}\rm{R. Horsley and P.~E.~L. Rakow}}

\author{W.~Bietenholz$^a$,
        V.~Bornyakov$^b$,
        M.~G\"ockeler$^c$,
        T. Hemmert$^c$,
        R. Horsley\thanks{Joint speakers}\footnotemark[1]$\,\,\,^{\,d}$,
        W.~G. Lockhart$^e$,
        Y.~Nakamura$^{c\,f}$,
        H.~Perlt$^g$,
        D.~Pleiter$^h$,
        P.~E.~L.~Rakow\footnotemark[1]$\,\,\,^e$,
        A.~Sch\"afer$^c$,
        G.~Schierholz$^{c\,i}$,
        A.~Schiller$^g$,
        T. Streuer$^{c}$,
        H.~St\"uben$^j$,
        F. Winter$^c$
        and J.~M.~Zanotti$^d$ \\
        \llap{$^a$} Instituto de Ciencias Nucleares,
                    Universidad Aut\'{o}noma de M\'{e}xico,
                    A.P. 70-543, C.P. 04510 Distrito Federal, Mexico \\
        \llap{$^b$} Institute for High Energy Physics,
                    142281 Protovino, Russia and \\
                    Institute of Theoretical and Experimental Physics,
                    117259 Moscow, Russia \\
        \llap{$^c$} Institut f\"ur Theoretische Physik,
                    Universit\"at Regensburg,
                    93040 Regensburg, Germany \\
        \llap{$^d$} School of Physics and Astronomy,
                    University of Edinburgh,
                    Edinburgh EH9 3JZ, UK \\
        \llap{$^e$} Theoretical Physics Division,
                    Department of Mathematical Sciences,
                    University of Liverpool,
                    Liverpool L69 3BX, UK \\
        \llap{$^f$} Center for Computational Sciences,
                    University of Tsukuba, Tsukuba,
                    Ibaraki 305-8577, Japan\thanks{Present address} \\
        \llap{$^g$} Institut f\"ur Theoretische Physik,
                    Universit\"at Leipzig,
                    04109 Leipzig, Germany \\
        \llap{$^h$} Deutsches Elektronen-Synchrotron DESY,
                    15738 Zeuthen, Germany \\
        \llap{$^i$} Deutsches Elektronen-Synchrotron DESY,
                    22603 Hamburg, Germany \\
        \llap{$^j$} Konrad-Zuse-Zentrum f\"ur Informationstechnik Berlin,
                    14195 Berlin, Germany  \\
        E-mail: \email{rhorsley@ph.ed.ac.uk, rakow@amtp.liv.ac.uk} }

\author{QCDSF--UKQCD Collaboration}

\abstract{QCD lattice simulations with 2+1 flavours typically start at
          rather large up-down and strange quark masses and extrapolate
          first the strange quark mass to its physical value and then
          the up-down quark mass. An alternative method of tuning the
          quark masses is discussed here in which the singlet quark mass
          is kept fixed, which ensures that the kaon always has mass less
          than the physical kaon mass. Using group theory the possible
          quark mass polynomials for a Taylor expansion about the
          flavour symmetric line are found, which enables highly
          constrained fits to be used in the extrapolation of hadrons 
          to the physical pion mass. Numerical results confirm the
          usefulness of this expansion and an extrapolation to the physical
          pion mass gives hadron mass values to within a few percent
          of their experimental values.}

\FullConference{The XXVIII International Symposium on Lattice Field Theory,
                Lattice2010  \\
		June 14-19, 2010  \\
		Villasimius, Italy}

\begin{document}


\section{Introduction} 


The QCD interaction is flavour-blind. Neglecting 
electromagnetic and weak interactions, the only difference between
quark flavours comes from the quark mass matrix. We investigate here
how flavour-blindness constrains hadron masses after flavour
$SU(3)$ is broken by the mass difference between the strange and 
light quarks, to help us extrapolate $2+1$ flavour lattice data 
to the physical point.

We have our best theoretical understanding when all $3$ quark flavours
have the same masses (because we can use the full power of flavour
$SU(3)$); nature presents us with just one instance of the theory, 
with $ m_s^{\R}/m_l^{\R} \approx 25$. We are interested in interpolating
between these two cases.  

\begin{figure}[htb]
   \vspace*{0.15in}
   \begin{center}
      \includegraphics[width=5.5cm]{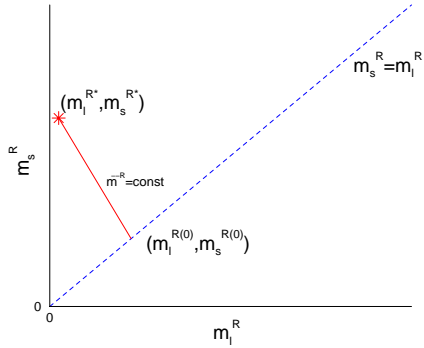}
   \end{center} 
   \caption{Sketch of the path (red, solid line) in the $(m_l^{\R}, m_s^{\R})$
            plane to the physical point $(m_l^{\R\,*}, m_s^{\R\,*})$.}
   \label{sketch_mlR_msR+path} 
\end{figure}

We consider possible behaviours near the symmetric point, and find
that flavour blindness is particularly helpful if we approach 
the physical point, $(m_l^{\R\,*}, m_s^{\R\,*})$, along a path
in the $(m_l^{\R}, m_s^{\R})$ plane starting at a point
on the $SU(3)$ flavour symmetric line ($m_l^{\R} = m_s^{\R}$)
and holding the sum of the quark masses
$m_u^{\R} + m_d^{\R} + m_s^{\R} \equiv 2m_l^{\R} + m_s^{\R}$ constant,
\cite{bientenholz10a}, as sketched in Fig.~\ref{sketch_mlR_msR+path}.


\section{Theory} 


Our strategy is to start from a point with all $3$ sea quark masses equal, 
\begin{eqnarray}
    m_u^{\R} = m_d^{\R} = m_s^{\R} \equiv m_0^{\R} \,,
\end{eqnarray}
and extrapolate towards the physical point, 
keeping the average sea quark mass
\begin{eqnarray}
   \overline{m}^{\R} \equiv \third (m_u^{\R} + m_d^{\R} + m_s^{\R}) 
\end{eqnarray}
constant. For this trajectory to reach the physical point
we have to start at a point where $m_0^{\R} \approx \third m_s^{\R\,*}$.  
As we approach the physical point, the $u$ and $d$ quarks
become lighter, but the $s$ becomes heavier. 
Pions are decreasing in mass, but $K$ and $\eta$ 
increase in mass as we approach the physical point.

We introduce the notation 
\begin{eqnarray} 
   \delta m_q^{\R}
         &\equiv& m_q^{\R} - \overline{m}^{\R} \,,  \qquad q = u, d, s \,,
 \end{eqnarray}   
and later use a similar notation for bare quark masses.
(We will be mainly interested in the $2+1$ flavour case, with
$m_u^{\R} = m_d^{\R} \equiv m_l^{\R}$.) With this notation,
the quark mass matrix is 
\begin{eqnarray}
   \cal M &=&  \left( \begin{array}{ccc}
                         m_u^{\R} & 0       & 0        \\
                         0       & m_d^{\R} & 0        \\
                         0       & 0        & m_s^{\R} \\ 
                      \end{array}
               \right)
                                                           \nonumber \\ 
          &=& \overline{m}^{\R}
              \left(  \begin{array}{ccc}
                         1 & 0 & 0 \\
                         0 & 1 & 0 \\
                         0 & 0 & 1 \\
                      \end{array}
              \right)
              + \half (\delta m_u^{\R} - \delta m_d^{\R})
              \left(  \begin{array}{ccc}
                         1 & 0  & 0  \\
                         0 & -1 & 0  \\
                         0 & 0  & 0  \\
                      \end{array}
              \right)
              + \half \delta m_s^{\R}
              \left(  \begin{array}{ccc}
                        -1 & 0 & 0 \\
                         0 &-1 & 0 \\
                         0 & 0 & 2 \\
                      \end{array}
              \right) \,.
\label{massmat} 
\end{eqnarray}  
The mass matrix ${\cal M}$  has a singlet part (proportional to $I$)
and an octet part, proportional to $\lambda_3$, $\lambda_8$. 
In the $2+1$ case $\muh^{\R} = \mdh^{\R}$ and isospin is a good symmetry. 
We argue that the theoretically cleanest way to
approach the physical point is to keep the singlet part
of ${\cal M}$ constant, and vary only the non-singlet parts.
One technical advantage of this strategy is that it simplifies
the quark mass renormalisation. In the case of clover/Wilson fermions,
the singlet and non-singlet parts of the mass matrix will renormalise
with different renormalisation constants \cite{gockeler04a}
\begin{eqnarray}
   m_q^R = Z_m^{NS}(m_q + \alpha_Z\overline{m}) \,, \qquad
   \alpha_Z = { Z_m^{\Si} - Z_m^{\NS} \over Z_m^{\NS} } \,,
\label{mr2mbare}
\end{eqnarray}
where $\alpha_Z$ represents the fractional difference between
the renormalisation constants. (Numerically this factor is
$\sim O(1)$, and is thus non-negligible. Of course,
for chiral fermions $\alpha_Z = 0$.) This gives
\begin{eqnarray}
   \overline{m}^{\R} = Z_m^{\NS}(1 + \alpha_Z)\overline{m} \,,
\end{eqnarray}
and so by keeping the singlet mass constant we avoid the need
to use two different $Z$s. This means that even for clover actions
it does not matter whether we keep the bare or renormalised average sea
quark mass constant (so we shall drop the $^{\R}$ superscript in the
following considerations).

\begin{figure}[htb]
   \vspace*{0.15in}
   \begin{center}
      \includegraphics[width=3cm,angle=270]{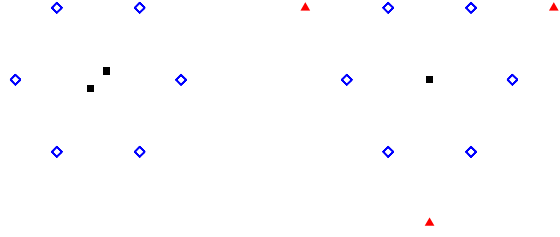}
    \end{center} 
\caption{The behaviour of the octet and decuplet under the
         permutation group $S_3$.
         The colours denote sets of particles which are invariant
         under permutations of the quark flavours.}
\label{permset}
\end{figure}

An important advantage of our strategy is that it strongly constrains
the possible mass dependence of physical quantities, and so simplifies
the extrapolation towards the physical point. 
Consider a flavour singlet quantity (for example the scale $r_0$, 
or the plaquette $P$) at the symmetric point $(m_0, m_0, m_0)$. 
If we make small changes in the quark masses, symmetry requires 
\begin{eqnarray}
   {\partial r_0 \over \partial m_u}
      = {\partial r_0 \over \partial m_d}
      = {\partial r_0 \over \partial m_s} \,.
\end{eqnarray}
If we keep $m_u + m_d + m_s$ constant, 
$d m_s = - d m_u - d m_d = -2 d m_l$ so 
\begin{eqnarray}
   d r_0 = d m_u \frac{\partial r_0}{\partial m_u}
           + d m_d \frac{\partial r_0}{\partial m_d}
           + d m_s \frac{\partial r_0}{\partial m_s} = 0 \,.
\label{symarg} 
\end{eqnarray}
The effect of making the strange
quark heavier exactly cancels the effect of making the light quarks
lighter, so we know that $r_0$ must have be stationary at the
symmetrical point. This makes extrapolations towards the physical 
point much easier, especially since we find that in practice 
quadratic terms in the quark mass expansion are very small. 
Any permutation of the quarks, such as an interchange 
$ u \leftrightarrow s$, or a cyclic permutation
$ u \to d \to s \to u $ doesn't change the physics,
it just renames the quarks. Any quantity unchanged by all
permutations will be flat at the symmetric point, like $r_0$. 
We can also construct permutation-symmetric combinations of hadrons. 
For example, for the decuplet, any permutation of the quark labels
will leave the $\Sigma^{0*}$, ($uds$) unchanged, so the $\Sigma^{0*}$
is shown by a single black point in Fig.~\ref{permset}.
On the other hand, a permutation (such as $u \to d \to s$)
can change a $\Delta^{++}(uuu)$ into a $\Delta^-(ddd)$
or (if repeated) into an $\Omega^-(sss)$, so these three
particles form a set of baryons which is closed under
quark permutations, and are all given the same colour (red)
in Fig.~\ref{permset}. Finally the $6$ baryons consisting 
of two quarks of one flavour, and one quark of a different flavour, 
form an invariant set, shown in blue in Fig.~\ref{permset}.
If we sum the masses in any of these sets, we get a flavour-symmetric 
quantity, which will obey the same argument we gave in 
eq.~(\ref{symarg}) for the quark mass (in)dependence of
the scale $r_0$. We therefore expect that the $\Sigma^{0*}$
mass must be flat at the symmetric point, and furthermore
that the combinations $(M_{\Delta^{++}} + M_{\Delta^-} + M_\Omega)$ 
and $(M_{\Delta^{+}} + M_{\Delta^0} + M_{\Sigma^{*+}}+  M_{\Sigma^{*-}} 
+ M_{\Xi^{*0}} + M_{\Xi^{*-}} )$ will also be flat. 
Technically these symmetrical combinations are in the $A_1$
singlet representation of the permutation group%
\footnote{$S_3$ has the same symmetry group as that of an equilateral
triangle, $C_{3v}$. This group has $3$ irreducible representations,
\cite{atkins70a}, two different singlets, $A_1$ and $A_2$ and a
doublet $E$, with elements $E^+$, $E^-$.}
$S_3$. We list some of these invariant mass
combinations in Table~{\ref{perminv}}.
\begin{table}[htb] 
   \begin{center} 
      \begin{tabular}{|l|c|c|}
      \hline
         Decuplet & $2 M_\Delta + M_\Omega$                   & red   \\
         baryons  & $2( M_\Delta + M_{\Sigma^*} + M_{\Xi^*} )$ & blue   \\
                  & $ M_{\Sigma^*}$                           & black \\
      \hline 
         Octet    & $2 ( M_N + M_\Sigma + M_\Xi )$            & blue  \\
         baryons  & $M_\Sigma + M_\Lambda$                    & black \\
      \hline
         Pseudoscalar
                  & $4 M_K^2 + 2 M_\pi^2$                     & blue  \\
         mesons   & $M^2_\pi + M^2_{\eta_8}$                  & black  \\
      \hline 
         Vector   & $4 M_{K^*} + 2 M_\rho$                    & blue   \\
         mesons   & $2M_\rho + M_{\phi_s}$                    & black  \\
      \hline 
      \end{tabular}
   \end{center}
\caption{Permutation invariant mass combinations,
         see Fig.~\protect\ref{permset}. $\phi_s$ is a
         fictitious $s \overline{s}$ particle; $\eta_8$
         a pure octet meson.
         The colours in the third column correspond to
         Fig.~\protect\ref{permset}.}
\label{perminv}
\end{table} 
We can use the singlet combinations from this table
to locate the starting point of our path to physics
by fixing a dimensionless ratio such as
\begin{eqnarray}
  {X_\pi^2 \over X_N^2} = \mbox{physical value} \,,
\label{XpioXN}
\end{eqnarray}
where $X_\pi^2 = \third(2 M_K^2 + M_\pi^2)$ and
$X_N = \third(M_N + M_\Sigma + M_\Xi)$.
The permutation group $S_3$ yields a lot of useful relationships,
but cannot capture the entire structure. For example, there is no way to
make a connection between the $\Delta^{++}(uuu)$
and the $\Delta^+(uud)$ by permuting quarks. 
To go further, we need to classify physical quantities by $SU(3)$
and the permutation group $S_3$ (which is a subgroup of $SU(3)$).

Let us first consider linear terms in $\delta m_q$.
These are given in Table~\ref{quad}.
\begin{table}[htb]
\begin{center}
   \begin{tabular} {|c|c|c|cccc|}
   \hline
      Polynomial  & & $S_3$ &  \multicolumn{4}{|c|}{$SU(3)$}  \\ 
   \hline
   \hline
      $ 1  $ & \checkmark & $A_1$&   1 &   &   &              \\
   \hline
   \hline
      $(\mbar - m_0)  $ & & $A_1$&   1 &   &   &              \\
   \hline
      $\msh$  & \checkmark &$E^+$&     & 8 &   &              \\
      $ (\muh - \mdh)  $ & \checkmark & $E^-$&   & 8 &  &     \\
   \hline
   \hline
      $  (\mbar - m_0)^2  $ &  & $A_1$&   1 &   &  &          \\
      $ (\mbar - m_0) \msh$ &  &$E^+$ &     & 8 &  &          \\
      $ (\mbar - m_0)(\muh - \mdh)$ & &$E^-$&   & 8 &  &      \\
   \hline
      $  \muh^2 + \mdh^2 +\msh^2 $  & \checkmark & $A_1$ &  1 && 27 & \\
      $ 3 \msh^2 - (\muh - \mdh)^2$  & \checkmark & $E^+$ & & 8 & 27 & \\
      $ \msh (\mdh - \muh)$  & \checkmark & $E^-$ &  & 8 & 27 & \\
   \hline
   \hline
\end{tabular}
\end{center}
\caption{All the quark-mass polynomials up to $O(m_q^2)$, classified by
         symmetry properties. A tick (\checkmark) marks the polynomials
         relevant on a constant $\mbar$ surface. These polynomials are
         plotted in Fig.~\protect\ref{triplot}.}
\label{quad}
\end{table}
Since we are keeping $\mbar$ constant, we are only changing the 
octet part of the mass matrix in eq.~(\ref{massmat}). Therefore, to first 
order in the mass change, only octet quantities can be effected.
$SU(3)$ singlets have no linear dependence on the quark mass, 
as we have already seen by the symmetry argument eq.~(\ref{symarg}), 
but we now see that all quantities in $SU(3)$ multiplets higher
than the octet cannot have linear terms. This provides a
constraint on the hadron masses within a multiplet and leads
to the Gell-Mann Okubo mass relations~\cite{gell-mann62a}. 

In the $2+1$ limit the decuplet baryons have $4$ different
masses (for the $\Delta$, $\Sigma^*$, $\Xi^*$, and $\Omega$),
but there is only one slope parameter in the linear mass formula.
Similarly, for the octet baryons there are $4$ distinct masses,
$(N,\Lambda, \Sigma, \Xi)$, but only $2$ slopes;
and for octet mesons, one slope parameter for three mesons,
($\pi, K, \eta$). Mesons have fewer slope parameters than
baryons because of constraints due to charge conjugation.
In the meson octet the $K$ and $\overline{K}$ must have
the same mass, but there is no reason why the $N$ and $\Xi$,
(which occupy the corresponding places in the baryon
octet), should have equal masses once flavour $SU(3)$ is broken. 

\begin{figure}[htb] 
   \begin{center} 
      \includegraphics[width=4.50cm,angle=270]{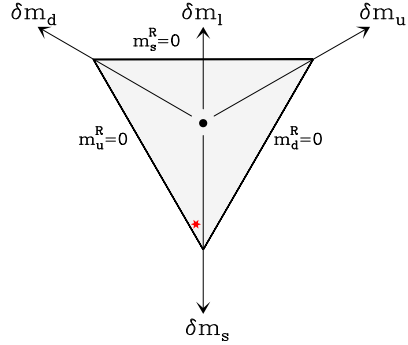}
   \end{center} 
\caption{The allowed quark mass region on the $\overline{m} = \mbox{const.}$
         surface is an equilateral triangle. The black point at the center
         is the symmetric point, the red star is the physical point.
         $2+1$ simulations lie on the vertical symmetry axis.
         The physical point is slightly off the $2+1$ axis because
         $m_d > m_u$.}
\label{trisk}
\end{figure}

\begin{figure}[htb] 
   \begin{center} 
      \includegraphics[width=9.50cm]{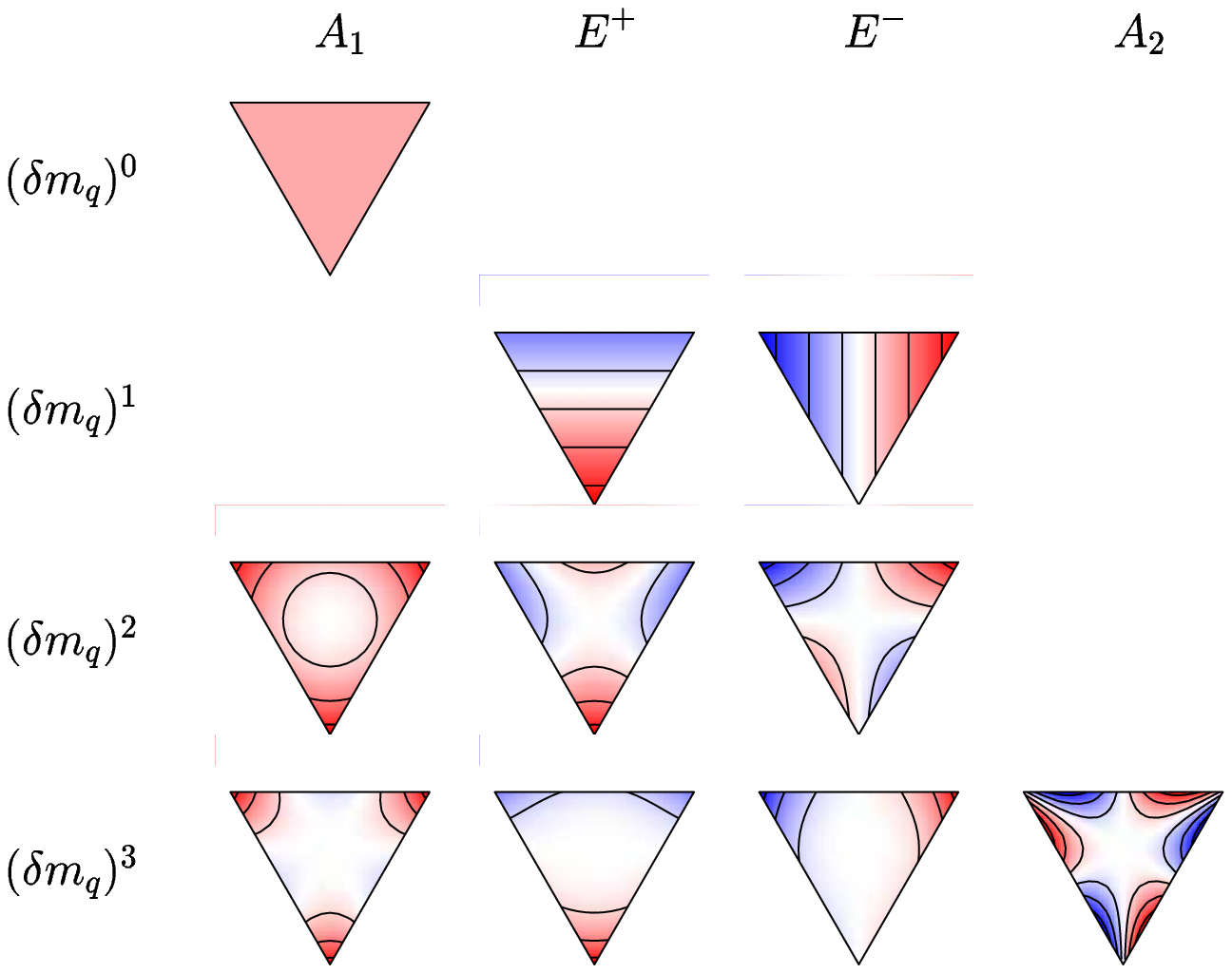}
  \end{center} 
\caption{Contour plots of the polynomials relevant for the
         constant $\mbar$ Taylor expansion, see Tables~\protect\ref{quad},
         \protect\ref{cubic}. A red(dish) colour denotes a positive number
         while a blue(ish) colour indicates a negative number.
         If $m_u=m_d$ (the $2+1$ case), only the polynomials in the
         $A_1$ and $E^+$ columns contribute.}
\label{triplot}
\end{figure}

When we proceed to quadratic polynomials we can construct 
polynomials which transform like mixtures of the $1$, $8$ and $27$ 
multiplets of $SU(3)$, Table~\ref{quad}.
This covers all the structures that can arise in the octet mass matrix,
but the decuplet mass matrix can include terms with the symmetries
$10$, ${\overline{10}},$ and $64$, which first occur when we look
at cubic polynomials in the quark masses, Table~\ref{cubic}. 
\begin{table}[htb]
\begin{center}
   \begin{tabular} {|c|c|c|cccccc|}
      \hline
      Polynomial  & & $S_3$ &  \multicolumn{6}{|c|}{$SU(3)$}  \\
      \hline
      \hline
      $(\mbar - m_0)^3$     &  &$A_1$& 1 &   &  &  &    &      \\
      $(\mbar - m_0)^2 \msh$&  &$E^+$&   & 8 &  &  &    &      \\
      $(\mbar - m_0)^2 (\muh - \mdh)$
                            &  &$E^-$&   & 8 &  &  &    &      \\
      $(\mbar - m_0)(\muh^2 + \mdh^2 +\msh^2 )$
                            &  &$A_1$& 1 &   &  &  & 27 &      \\
      $(\mbar - m_0)\left[3 \msh^2 - (\muh - \mdh)^2 \right]$
                            &  &$E^+$&   & 8 &  &  & 27 &      \\
      $(\mbar - m_0)\msh (\mdh - \muh)$ 
                            &  &$E^-$&   & 8 &  &  & 27 &      \\
      \hline
      $\muh \mdh \msh $     &
                             \checkmark&$A_1$& 1 &   &  &  & 27 & 64 \\
      $\msh (\muh^2 + \mdh^2 +\msh^2 )$
                            &\checkmark&$E^+$&   & 8 &  &  & 27 & 64 \\
      $(\muh - \mdh) (\muh^2 + \mdh^2 +\msh^2 )$
                            &\checkmark&$E^-$&   & 8 &  &  & 27 & 64 \\
      $(\msh - \muh)(\msh-\mdh)(\muh-\mdh)$
                            &\checkmark&$A_2$&   &   &10& $\overline{10}$
                                                           &    & 64 \\
      \hline
      \hline
   \end{tabular}
\end{center}
\caption{The cubic quark-mass polynomials, classified by
         symmetry properties. A tick (\checkmark) marks the polynomials
         relevant on a constant $\mbar$ surface. These polynomials are
         plotted in Fig.~\protect\ref{triplot}.}
\label{cubic}
\end{table}
The allowed quark mass region on the $\overline{m} = \mbox{const.}$
surface is an equilateral triangle, as shown in Fig.~\ref{trisk}.
Plotting the polynomials of Tables~\ref{quad}, \ref{cubic} then gives
the figures in Fig.~\ref{triplot}, where the colour coding indicates
whether the polynomial is positive (red) or negative (blue).

We can see how well this works in practice by looking at the physical 
masses of the decuplet baryons, 
\begin{eqnarray}
   4 M_\Delta + 3 M_{\Sigma^*} + 2 M_{\Xi^*} + M_{\Omega}
      &=& 13.821 {\rm \ GeV}
          \qquad \quad \ {\rm singlet}
          \quad \propto (\delta m_l)^0 \label{num1}           \nonumber  \\
   - 2 M_\Delta \qquad\ \quad  + M_{\Xi^*} + M_{\Omega}
      &=& \ 0.742 {\rm \ GeV} 
          \qquad \quad \ {\rm octet}
          \qquad \propto (\delta m_l)^1 \label{num8}          \nonumber  \\
   4 M_\Delta - 5 M_{\Sigma^*} - 2 M_{\Xi^*} + 3 M_{\Omega}
      &=& -0.044  {\rm \  GeV} 
          \qquad \quad {\rm 27plet}
          \qquad \propto (\delta m_l)^2 \label{num27}         \nonumber  \\
   - M_\Delta + 3 M_{\Sigma^*} - 3 M_{\Xi^*} +  M_{\Omega}
      &=& -0.006 {\rm \  GeV}  \qquad \quad {\rm 64plet} \qquad 
          \propto  (\delta m_l)^3\qquad \hbox{}
\label{num64}
\end{eqnarray}
When we form combinations with particular $SU(3)$ symmetries we 
see a strong hierarchy, which suggests a short Taylor series may
work well all the way from  symmetry point $(m_0, m_0, m_0)$
to the physical point. This gives the constrained fit formulae
\begin{eqnarray}
   M_\pi^2      &=& M_0^2 + 2\alpha\delta m_l 
                         + (\beta_0 + 2\beta_1)\delta m_l^2
                                                             \nonumber \\
   M_K^2        &=& M_0^2 - \alpha\delta m_l 
                         + (\beta_0 + 5\beta_1 + 9\beta_2)\delta m_l^2
                                                             \nonumber \\
   M_{\eta_s}^2  &=& M_0^2 - 4\alpha\delta m_l 
                         + (\beta_0 + 8\beta_1)\delta m_l^2 \,,
\label{fit_mpsO}
\end{eqnarray}
\begin{eqnarray}
   M_\rho       &=& M_0 + 2\alpha\delta m_l 
                         + (\beta_0 + 2\beta_1)\delta m_l^2
                                                             \nonumber \\
   M_{K^*}      &=& M_0 - \alpha\delta m_l 
                         + (\beta_0 + 5\beta_1 + 9\beta_2)\delta m_l^2
                                                             \nonumber \\
   M_{\phi_s}   &=& M_0 - 4\alpha\delta m_l 
                         + (\beta_0 + 8\beta_1)\delta m_l^2 \,,
\label{fit_mvO}
\end{eqnarray}
\begin{eqnarray}
   M_N      &=& M_0 + 3A_1\delta m_l 
                    + (B_0+3B_1)\delta m_l^2
                                                             \nonumber \\
   M_\Lambda &=& M_0  + 3A_2\delta m_l 
                    + (B_0+6B_1-3B_2+9B_4)\delta m_l^2
                                                             \nonumber \\
   M_\Sigma &=& M_0  - 3A_2\delta m_l 
                    + (B_0+6B_1+3B_2+9B_3)\delta m_l^2
                                                             \nonumber \\
   M_\Xi    &=& M_0 - 3(A_1-A_2)\delta m_l 
                    + (B_0+9B_1-3B_2+9B_3)\delta m_l^2 \,,
\label{fit_mNO}
\end{eqnarray}
\begin{eqnarray}
   M_\Delta     &=& M_0 + 3A\delta m_l 
                       + (B_0 + 3B_1)\delta m_l^2
                                                             \nonumber \\
   M_{\Sigma^*} &=& M_0  + 0
                       + (B_0 + 6B_1 + 9B_2)\delta m_l^2
                                                             \nonumber \\
   M_{\Xi^*}    &=& M_0 - 3A\delta m_l 
                       + (B_0 + 9B_1 + 9B_2)\delta m_l^2
                                                             \nonumber \\
   M_\Omega    &=& M_0 - 6A\delta m_l 
                       + (B_0 + 12B_1)\delta m_l^2 \,.
\label{fit_mDD}
\end{eqnarray}
While the linear terms are highly constrained, the quadratic
terms much less so; indeed only for the baryon decuplet is there
any constraint%
\footnote{The coefficients of the $\delta m_l^2$ term appear
complicated; indeed there seem to be too many for the nucleon octet,
eq~(\ref{fit_mNO}). However although not discussed here the $SU(3)$
flavour symmetry breaking expansion can be extended to different
valence quarks than sea quarks or `partially quenching'.
In this case the $m_l$, $m_s$ sea quarks remain constrained
by $\overline{m} = \mbox{const.}$, but the valence quarks
denoted by $\mu_l$, $\mu_s$ are unconstrained. With
$\delta\mu_q = \mu_q - \overline{m}$, the nucleon octet,
eq.~(\ref{fit_mNO}) becomes
\begin{eqnarray}
   M_N &=& M_0 + 3A_1\delta\mu_l + B_0\delta m_l^2 + 3B_1\delta\mu_l^2
                                                             \nonumber \\
   M_\Lambda
       &=& M_0 + A_1(2\delta\mu_l + \delta\mu_s) 
               - A_2(\delta\mu_s - \delta\mu_l)
               + B_0\delta m_l^2
               + B_1(2\delta\mu_l^2 + \delta\mu_s^2)
               - B_2(\delta\mu_s^2 - \delta\mu_l^2)
               + B_4(\delta\mu_s - \delta\mu_l)^2
                                                             \nonumber \\
   M_\Sigma
       &=& M_0 + A_1(2\delta\mu_l + \delta\mu_s) 
               + A_2(\delta\mu_s - \delta\mu_l)
               + B_0\delta m_l^2
               + B_1(2\delta\mu_l^2 + \delta\mu_s^2)
               + B_2(\delta\mu_s^2 - \delta\mu_l^2)
               + B_3(\delta\mu_s - \delta\mu_l)^2
                                                             \nonumber \\
   M_\Xi
       &=& M_0 + A_1(2\delta\mu_l + \delta\mu_s) 
               - A_2(\delta\mu_s - \delta\mu_l)
               + B_0\delta m_l^2
               + B_1(\delta\mu_l^2 + 2\delta\mu_s^2)
               - B_2(\delta\mu_s^2 - \delta\mu_l^2)
               + B_3(\delta\mu_s - \delta\mu_l)^2 \,.
                                                             \nonumber
\end{eqnarray}
When $\mu \to m$ (i.e.\ return to the `unitary line') then these
results collapse to the previous results of eq.~(\ref{fit_mNO}).
Similar results hold for eqs.~(\ref{fit_mpsO}), (\ref{fit_mvO})
and (\ref{fit_mDD}). We can also show that on this trajectory
the errors of the partially quenched approximation are much smaller
than on other trajectories.}
. Note also that for the pseudoscalar octet,
$M_{\eta_s}$ is a fictitious $s\overline{s}$ particle
(due to non-perfect $\eta$-$\eta^\prime$ mixing),
while for the vector octet due to near perfect mixing
between the $\phi$ and $\omega$ the $M_\phi$ is (almost) a perfect
$s\overline{s}$ state, so that $M_{\phi_s} \approx M_\phi$.
However even these fictitious particles still contain useful
information, due to the constrained fit.

As eqs.~(\ref{fit_mpsO})-(\ref{fit_mDD}) have been derived using
only group theoretic arguments, then they will be valid
for results derived on any lattice volume (though the coefficients
are still functions of the volume). 

\begin{figure}[htb]
   \vspace*{0.15in}
   \begin{center}
      \includegraphics[width=5.0cm]{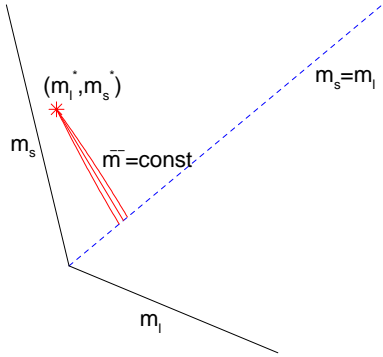}
   \end{center} 
   \caption{Sketch of some possible paths (red lines) in the $(m_l, m_s)$
            plane to the physical point $(m_l^{\R\,*}, m_s^{\R\,*})$.}
   \label{sketch_ml_ms_bLO+paths}
\end{figure}

What is the effect of higher order terms on our path in the
$(m_s, m_l)$ plane? While at lowest order it did not matter
whether we kept the quark mass constant or a particle mass
now it does. Practically it is easiest to keep the (bare)
quark mass fixed. Thus lines of constant $X_S$ are curved
though they do still have  to have the slope of $-2$
at the point where they cross the $SU(3)$ flavour symmetric line.
Now we have to specify more closely what we mean when we keep
$2m_K^2 + m_\pi^2$ constant, as different scale choices give
slightly different paths as sketched in Fig.~\ref{sketch_ml_ms_bLO+paths}.
Note that the physical domain $m_l^{\R}, m_s^{\R} \ge 0$ translates to
\begin{eqnarray}
   m_l \ge - { \third\alpha_Z \over (1 + \twothird\alpha_Z) } m_s
                                 \,, \qquad
   m_s &\ge& - { \twothird\alpha_Z \over (1 + \third\alpha_Z) } m_l \,,
\label{phy_dom_b}
\end{eqnarray}
leading to non-orthogonal axes and possibly negative bare quark mass.
(These features disappear for chiral fermions when $\alpha_Z = 0$.)
If we make different choices of the quantity we keep constant
at the experimentally measured physical value, for example
\begin{eqnarray}
   { X_\pi^2 \over X_N^2 }\,, \quad
   { X_\pi^2 \over X_\Delta^2 }\,, \quad
   { X_\pi^2 \over X_\rho^2 }\,, \quad
   { X_\pi^2 \over X_r^2 } \,,
\label{some_other_XpioXs}
\end{eqnarray}
where in addition to the previously defined singlet quantities we
also now have $X_\Delta = \third (2M_\Delta + M_\Omega)$, 
$X_\rho = \third (2M_{K^*} + M_\rho)$, $X_r = r_0^{-1}$
we get slightly different trajectories. The different trajectories
begin at slightly different points along the flavour $SU(3)$ symmetric
line. Initially they are all parallel with slope $-2$, but away from the
symmetry line they can curve, and will all meet at the physical 
point. (Numerically we shall later see that this seems to be a small
effect.)

Finally we mention the connection of this method to $\chi$PT.
For example for the pseudoscalar octet, using the results
in \cite{allton08a} and assuming their validity up to the
point on the flavour symmetric line, we find
\begin{eqnarray}
   M_0^2 &=& \overline{\chi}\,\left[
                1 - {16\overline{\chi}\over f_0^2}
                       \left( 3L_4 + L_5 -6L_6 -2L_8 \right)
                        + {\overline{\chi}\over 24\pi^2 f_0^2}
                          \ln {\overline{\chi}\over \Lambda_\chi^2}
                             \right]
                                                              \nonumber \\
  \alpha &=& Q_0 \,\left[
                1 - {16\overline{\chi}\over f_0^2}
                       \left( 3L_4 + 2L_5 -6L_6 -4L_8 \right)
                        + {\overline{\chi}\over 8\pi^2 f_0^2}
                          \ln {\overline{\chi}\over \Lambda_\chi^2}
                             \right]
                                                              \nonumber \\
  \beta_0&=& - {Q_0^2 \over 6\pi^2 f_0^2}
                                                              \nonumber \\
  \beta_1&=& {Q_0^2 \over f_0^2}\, \left[
                - 32\left(L_5 - 2L_8 \right)
                + {1\over 24\pi^2}
                    \left(7 + 4\ln {\overline{\chi}\over \Lambda_\chi^2} \right)
                                  \right]
                                                              \nonumber \\
  \beta_2&=& {Q_0^2 \over f_0^2}\, \left[
                 16\left(L_5 - 2L_8 \right)
                 - {1\over 24\pi^2}
                   \left(3 + 2\ln {\overline{\chi}\over \Lambda_\chi^2} \right)
                                  \right] \,,
\end{eqnarray}
with $Q_0 = B_0^{\R}Z_m^{\NS}$,
$\overline{\chi} = 2Q_0(1+\alpha_z)\overline{m}$,
where the $L$s are appropriate Low Energy Constants or LECs.
For clover fermions, as mentioned before we have to respect the
fact that singlet and non-singlet quark masses renormalise
differently which leads to a non-zero $\alpha_Z$.

We first note that when expanding the $\chi$-PT about a point on
the $SU(3)$ flavour symmetry line gives to leading order
only one parameter -- $\alpha$. (This means, in particular,
that flavour singlet combinations vanish to leading order, as
discussed previously.) Secondly, while we fit to $\alpha$
and $\beta_0$, $\beta_1$ and $\beta_2$, it will be difficult
to determine the individual LECs. The best we can probably
hope for are these combinations.


\section{The Lattice}


We use a clover action for $2+1$ flavours with a single iterated
mild stout smearing as described in \cite{cundy09a}. Also given
in this reference is a non-perturbative determination
of the improvement coefficient for the clover term, using the
Schr\"odinger Functional method. All the results given here will
be at $\beta = 5.50$, $c_{sw} = 2.65$. HMC and RHMC were used
for the $2$, $1$ fermion flavours respectively, \cite{nakamure10a},
to generate the gauge configurations.
The (bare) quark masses are defined as
\begin{eqnarray}
   am_q = {1 \over 2} 
            \left ({1\over \kappa_q} - {1\over \kappa_{sym;c}} \right) \,,
\end{eqnarray}
where vanishing of the quark mass along the $SU(3)$ flavour 
symmetric line determines $\kappa_{sym;c}$.
We then keep $\overline{m} = \mbox{const.} \equiv  m_0$ which gives
\begin{eqnarray}
   \kappa_s 
      = { 1 \over { {3 \over \kappa_{sym}^{(0)}} - {2 \over \kappa_l} } } \,.
\end{eqnarray}
So once we decide on $\kappa_l$ this then determines $\kappa_s$.
A series of runs along the $SU(3)$ flavour line determines the initial
point on this line: $\kappa_{sym}^{(0)}$ by looking when $X_\pi/X_S$,
$S = N$, $\Delta$, $\rho$ are equal to their physical values,
see eqs.~(\ref{XpioXN}), (\ref{some_other_XpioXs}). This would
also include $r$ if we have previously determined the physical value
of $r_0$.

\begin{figure}[htb]
   \vspace*{0.15in}
   \begin{center}
      \includegraphics[width=10.0cm]
             {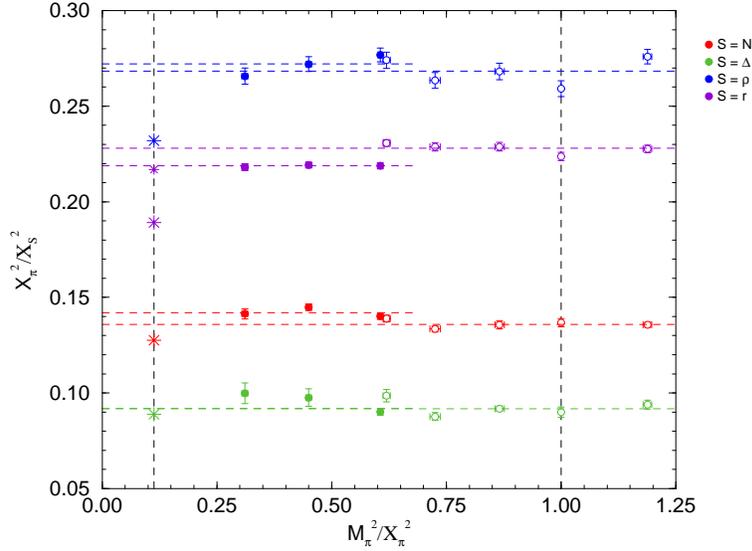}
   \end{center} 
   \caption{$X_\pi^2/X_S^2$ versus $M_\pi^2/X_\pi^2$ for
            $S = N$, $\Delta$, $\rho$, $r$ for $\kappa_{sym}^{(0)}=0.12090$.
            The dashed vertical line represents the physical value,
            while the dotted line gives the $SU(3)$ flavour symmetric point.
            Filled points are on $32^3\times 64$ lattices
            while open points are on $24^3\times 48$ sized lattice.
            Dashed horizontal lines represent constant fits
            to either the $32^3\times 64$ or $24^3\times 48$ results.
            The physical values are denoted by stars. For $r_0$
            the small star is the value obtained using
            $r_0 = 0.5\,\mbox{fm}$ while the large star uses
            $r_0 = 0.467\,\mbox{fm}$ (determined for $n_f=2$
            degenerate flavours \protect\cite{alikhan06a}).}
\label{b5p50_mps2oX2_2mpsK2+mps2o3oX2}
\end{figure}

\begin{figure}[htb]
   \vspace*{0.15in}
   \begin{center}
      \includegraphics[width=10.0cm]
             {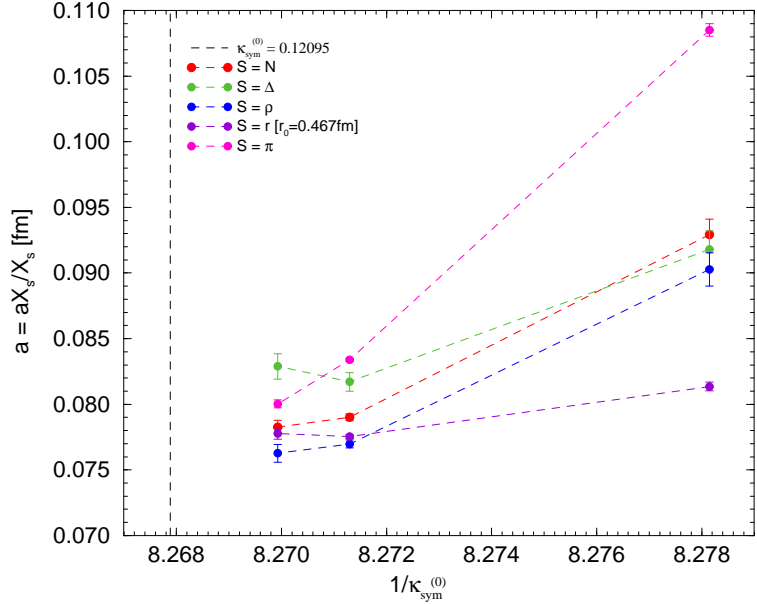}
   \end{center} 
   \caption{$aX_S / X_S$ against $1/\kappa^{(0)}_{sym}$ for
            $S = N$, $\Delta$, $\rho$, $r$, $\pi$
            and $\kappa^{(0)}_{sym} = 0.12080$, $0.12090$, $0.12092$.
            The vertical dashed line is at $\kappa^{(0)}_{sym} = 0.12095$.}
\label{b5p50_ookl_aXoXexpt}
\end{figure}

In Fig.~\ref{b5p50_mps2oX2_2mpsK2+mps2o3oX2} we plot $X_\pi^2/X_S^2$
for various $X_S$ (with $S = N$, $\Delta$, $\rho$, $r$). Also shown
are constant fits to the two volumes --- $24^3\times 48$ and
$32^3\times 64$. (The runs on $24^3\times 48$ lattices have $O(2000)$
trajectories, while the runs on $32^3\times 64$ lattices have
$O(1500)$ trajectories.) As mentioned before we first simulate at
various quark masses on the $SU(3)$ flavour symmetric line
(i.e.\ $m_\pi^2/X_\pi^2 = 1$ in Fig.~\ref{b5p50_mps2oX2_2mpsK2+mps2o3oX2}
where $\kappa_{sym}^{(0)} = 0.12090$) to determine $\kappa_{sym}^{(0)}$.
Note that simulations with a `light' strange quark mass and heavy
`light' quark mass are possible -- here the right most point.
(In this inverted strange world we would expect $p \to \Sigma$
or $\Lambda$ decays.) It is apparent that while some scales fluctuate others
(in particular $S = N$ and $r$) are very smooth. We take this
as a sign that singlet quantities are very flat and the fluctuations
are due to low statistics. Assuming that all ratios are constant,
(i.e.\ higher order effects are small, see eq.~(\ref{some_other_XpioXs}))
we would expect all the ratios to converge to their experimental
values. This is complicated because there are small finite size effects
present (this can again best be seen in the $S = N$ and $r$ data).
To investigate possible effects we are generating results on a 
variety of lattices, but for the present for the ratios we
simply take the largest volume available. In Fig.~\ref{b5p50_ookl_aXoXexpt}
we plot $aX_S / X_S$, $S = N$, $\Delta$, $\rho$, $r$, $\pi$
for the largest volume fitted results from 
Fig.~\ref{b5p50_mps2oX2_2mpsK2+mps2o3oX2} (together with smaller
data sets for $\kappa^{(0)}_{sym} = 0.12080$, $\kappa^{(0)}_{sym} = 0.12092$,
the latter data set is presently only partially generated). 
This ratio gives estimates for the lattice spacing $a$ for the various
scales. There appears to be convergence to a common scale
of $a \sim 0.078\,\mbox{fm}$. To investigate this point further
we are performing additional runs at $\kappa^{(0)}_{sym} = 0.12092$ and
$0.12095$.


\section{Spectrum results}

\begin{figure}[p]
   \begin{center}
      \includegraphics[width=10.0cm]
             {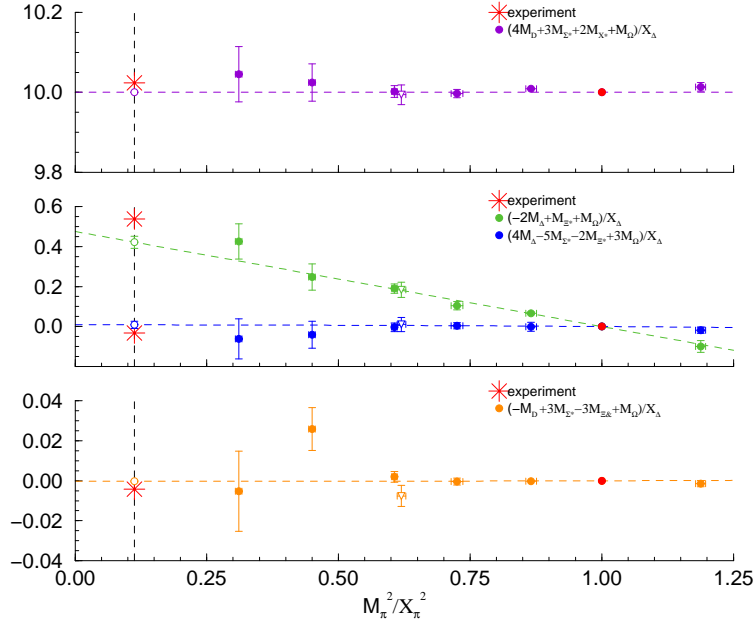}
   \end{center} 
   \caption{$(4M_\Delta + 3M_{\Sigma^*}+2M_{\Xi^*}+M_\Omega)/X_\Delta$,
            $(-2M_\Delta + M_{\Xi^*}+M_\Omega)/X_\Delta$,
            $(4M_\Delta - 5M_{\Sigma^*}-2M_{\Xi^*}+M_\Omega)/X_\Delta$
            and
            $(-M_\Delta + 3M_{\Sigma^*} - 3M_{\Xi^*} + M_\Omega)/X_\Delta$
            (filled circles) against
            $M_\pi^2/X_\pi^2$ together with a fit of constant,
            linear quadratic and cubic term in $\delta m_l$
            respectively. Extrapolated values are shown as opaque
            circles. Experimental values are denoted by stars.
            The opaque triangle is a run at the same
            $(\kappa_l,\kappa_s)$, but on a $24^3\times 48$ lattice
            rather than a $32^3\times 64$ lattice.}
\label{mps2o2mpsK2+mps2o3_su3NDsymo2mDelta+mOmDo3}
\end{figure}

We now consider the mass spectrum. First we check whether there is
a strong hierarchy due to the $SU(3)$ flavour symmetry as found
in eq.~(\ref{num64}).
In Fig.~\ref{mps2o2mpsK2+mps2o3_su3NDsymo2mDelta+mOmDo3}
we plot $(4M_\Delta + 3m_{\Sigma^*}+2M_{\Xi^*}+M_\Omega)/X_\Delta$,
$(-2M_\Delta + M_{\Xi^*}+M_\Omega)/X_\Delta$,
$(4M_\Delta - 5M_{\Sigma^*}-2M_{\Xi^*}+M_\Omega)/X_\Delta$
and $(-M_\Delta + 3M_{\Sigma^*} - 3M_{\Xi^*} + M_\Omega)/X_\Delta$
against $M_\pi^2/X_\pi^2$. Also shown are the experimental values.
There is reasonable agreement with these numbers. 
(See \cite{beane06a} for a similar investigation of octet baryons.)
It is also seen that as expected while
$(-2M_\Delta + M_{\Xi^*}+M_\Omega)/X_\Delta$
has a linear gradient in the pion mass, in the other fits
any gradient is negligible. To check for possible finite size
effects we also plot a run at the same $(\kappa_l,\kappa_s)$
but using $24^3\times 48$ lattice rather than $32^3\times 64$.
There is little difference and so it appears that considering
ratios of quantities within the same multiplet leads to (effective)
cancellation of finite size effects.

\begin{figure}[p]
   \begin{center}
      \includegraphics[width=10.0cm]
             {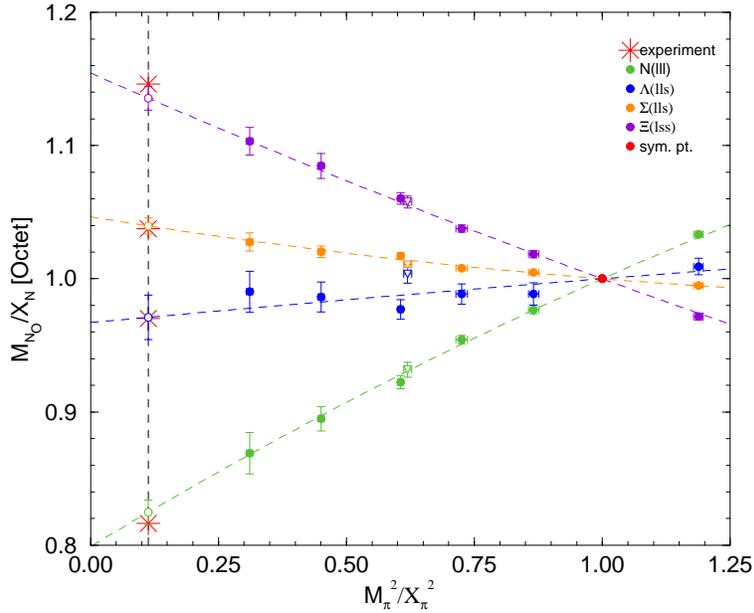}
   \end{center} 
   \caption{$M_{N_O}/ X_N$ ($N_O = N$, $\Lambda$, $\Sigma$, $\Xi)$)
            against $M_\pi^2/X_\pi^2$ together with the combined fit
            of eqs.~(\protect\ref{fit_mNO}), (\protect\ref{fit_mpsO})
            (the dashed lines). Experimental values are denoted
            by stars. The opaque triangle is a run at the same
            $(\kappa_l,\kappa_s)$, but on a $24^3\times 48$ lattice
            rather than a $32^3\times 64$ lattice.}
\label{b5p50_mpsO2o2mpsK2+mps2o3-jnt_mNOomNOpmSigOpmXiOo3-boot-jnt}
\end{figure}

\begin{figure}[p]
   \vspace*{0.15in}
   \begin{center}
      \includegraphics[width=10.0cm]
     {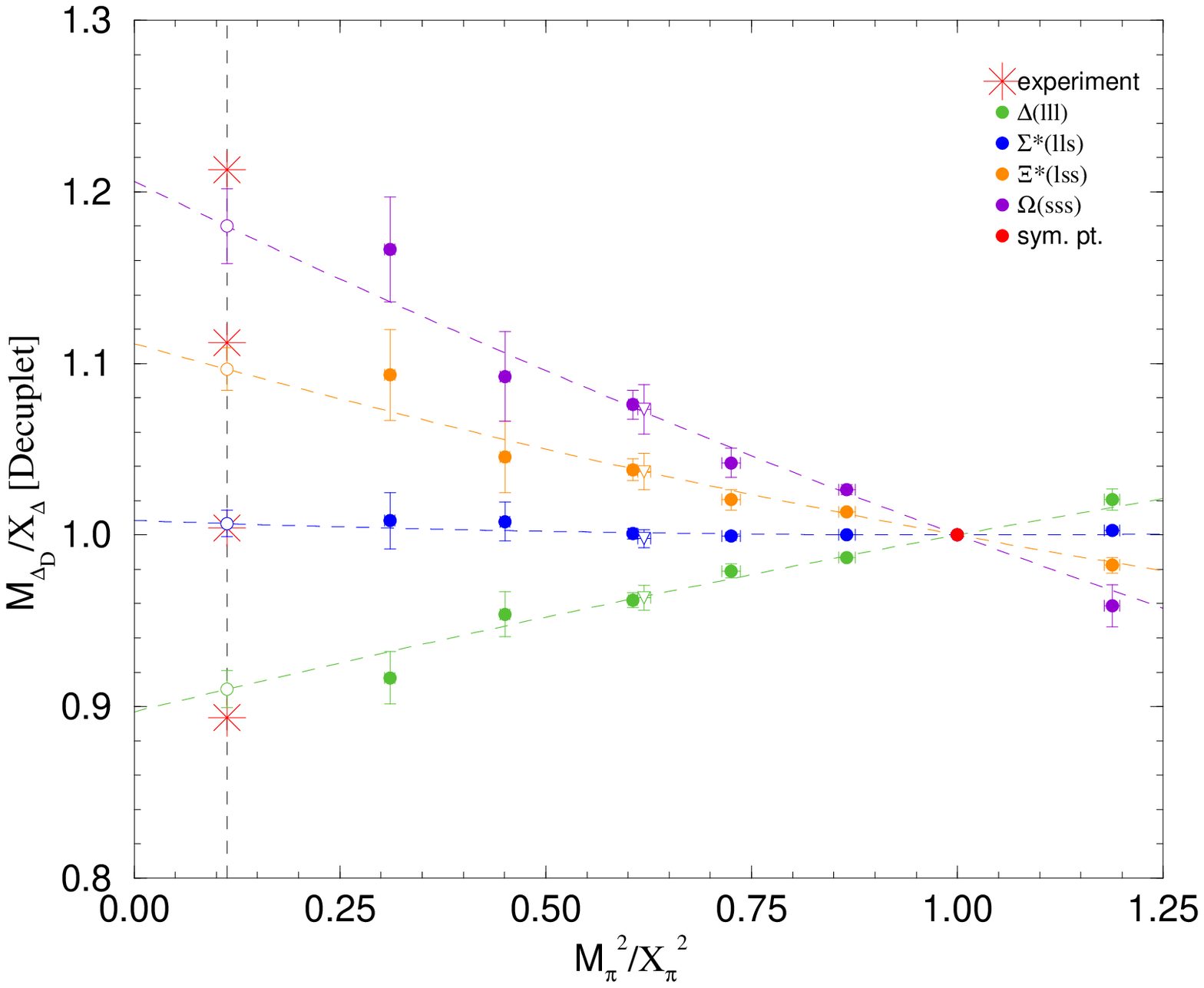}
   \end{center}
   \caption{$M_{\Delta_D}/ X_{\Delta}$
            ($\Delta_D = \Delta$, $\Sigma^*$, $\Xi^*$, $\Omega$)
            against $M_\pi^2/X_\pi^2$ together with the combined fit of
            eqs.~(\protect\ref{fit_mDD}), (\protect\ref{fit_mpsO}).
            Same notation as Fig.~\protect
            \ref{b5p50_mpsO2o2mpsK2+mps2o3-jnt_mNOomNOpmSigOpmXiOo3-boot-jnt}.}
\label{b5p50_mpsO2o2mpsK2+mps2o3-jnt_mDDo2mDeltaD+mOmDo3-boot-jnt}
\end{figure}

\begin{figure}[p]
   \vspace*{0.15in}
   \begin{center}
      \includegraphics[width=10.0cm]
             {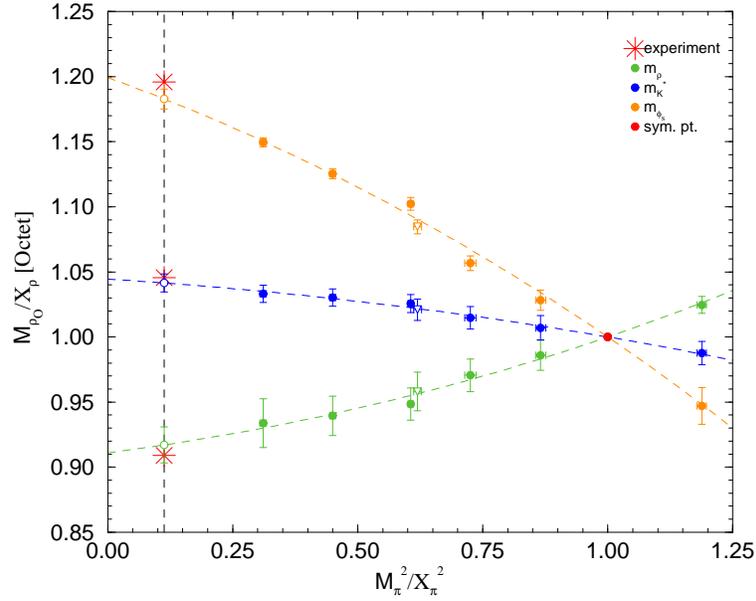}
   \end{center} 
   \caption{$M_{\rho_O}/ X_{\rho}$
            ($\rho_O = \rho$, $K^*$, $\phi_s$)
            against $M_\pi^2/X_\pi^2$ together with the combined fit of
            eqs.~(\protect\ref{fit_mvO}), (\protect\ref{fit_mpsO})
            (the dashed lines).
            Same notation as Fig.~\protect
            \ref{b5p50_mpsO2o2mpsK2+mps2o3-jnt_mNOomNOpmSigOpmXiOo3-boot-jnt}.}
\label{b5p50_psOX2-jnt_mvOoX-jnt}
\end{figure}

In Fig.~\ref{b5p50_mpsO2o2mpsK2+mps2o3-jnt_mNOomNOpmSigOpmXiOo3-boot-jnt}
we plot the nucleon octet $M_{N_O}/ X_N$ (for $N_O = N$, $\Lambda$,
$\Sigma$, $\Xi$) against $M_\pi^2/X_\pi^2$ for $\kappa_{sym} = 0.12090$.
A typical `fan' structure is seen with results radiating from the
common point on the symmetric line. Again, finite volume effects
tend to cancel in the ratio (normalising with the singlet quantity
from the same octet) and so both volumes have been used in the fit.
The combined fit uses eqs.~(\ref{fit_mNO}), (\ref{fit_mpsO}) with
the bare quark mass being an `internal' parameter. Note that
one point has a light strange quark and a heavy `light' quark. Similarly in
Figs.~\ref{b5p50_mpsO2o2mpsK2+mps2o3-jnt_mDDo2mDeltaD+mOmDo3-boot-jnt}
and \ref{b5p50_psOX2-jnt_mvOoX-jnt}
we plot the corresponding nucleon decuplet $M_{\Delta_D}/ X_\Delta$
and vector octet $M_{\rho_O}/ X_\rho$ against $M_\pi^2/X_\pi^2$
respectively. Although we have included quadratic terms in the
fit, there is really very little curvature in the results.
We find good agreement with the experimental results.


\section{Conclusions}


We have outlined a programme to systematically approach the
physical point starting from a point on the $SU(3)$ flavour symmetric
line. Exploratory results for the hadron mass spectrum show that
constrained linear extrapolations give accurate results for the
mass spectrum. We are also applying this method to the computation
of matrix elements, some initial results are given in
\cite{zanotti10a,winter10a}.


\section*{Acknowledgements}


The numerical calculations have been performed on the IBM
BlueGeneL at EPCC (Edinburgh, UK), the BlueGeneL and P at
NIC (J\"ulich, Germany), the SGI ICE 8200 at HLRN (Berlin-Hannover, Germany)
and the JSCC (Moscow, Russia). We thank all institutions.
The BlueGene codes were optimised using Bagel, \cite{boyle09a}.
This work has been supported in part by the EU grants
227431 (Hadron Physics2), 238353 (ITN STRONGnet)
and by the DFG under contract SFB/TR 55 (Hadron Physics from Lattice QCD).
JZ is supported by STFC grant ST/F009658/1.



\end{document}